\documentclass[preprint,12pt]{article}
\usepackage{graphicx} 
\usepackage{amsmath,amssymb}
\usepackage{amsfonts}
\usepackage{authblk}
\usepackage{todonotes}

\usepackage{comment}
\usepackage{subcaption}

\newcommand{\E}{\mathbb{E}}
\newcommand{\R}{\mathbb{R}}
\newcommand{\T}{\mathbb{T}}
\newcommand{\PP}{\mathbb{P}}

\newcommand{\eps}{\epsilon}
 
\title{Particle separation in a time-correlated Gaussian velocity field: a numerical study}

\author{Mario Maurelli\thanks{mario.maurelli@unipi.it} }
\author{Silvia Morlacchi\thanks{silvia.morlacchi@dm.unipi.it}}

\affil{%
  \small
  Dipartimento di Matematica, Universit\`a di Pisa\\
  Largo Pontecorvo 5, 56127 Pisa, Italy
}

\begin{document}
\maketitle

\begin{abstract}
    We numerically investigate the dispersion of passive tracer pairs in a two-dimensional, divergence-free, time-correlated Gaussian velocity field. The field is constructed from a sparse set of Fourier modes evolving as independent Ornstein–Uhlenbeck processes, with spatial regularity and scale-dependent temporal decorrelation comparable to those of a power-law Gaussian velocity ensemble.  We show that, for spatial roughness exponent \(\alpha=0.2\) and temporal decorrelation exponent \(\beta=0.5\), corresponding to the regime \(\alpha+\beta<1\), the extrapolated mean-square distance is compatible with particle separation (in mean-square) at positive time: two particles reach a mean-square positive distance at a finite time, no matter how close they are initially.
\end{abstract}

\textbf{Keywords}: Lagrangian particle dispersion, time-correlated Gaussian velocity fields, Kraichnan-type turbulence models

\textbf{MSC 2020 classification}: Primary: 76F25; Secondary: 76F55, 65C05

\section{Introduction}
\label{sec:introduction}

A long-standing problem in turbulence is the study of the relative motion of passive tracers. Its origin can be traced back to the work of Richardson \cite{richardson1926atmospheric}, predicting superdiffusive pair dispersion $\langle |X_t-Y_t|^2\rangle\propto t^3$, where $X$ and $Y$ are two tracer trajectories. This prediction raises the question of whether particles starting from distinct, increasingly close initial positions can separate at positive times, namely whether $\limsup_{|X_0-Y_0|\to 0}|X_t-Y_t|$ is positive at positive time $t$ (what we refer to as particle separation) or not. From the mathematical viewpoint, this problem is related to the (non)uniqueness and (in)stability for the ODE
\begin{equation}
    \dot X_t=v(t,X_t),
    \label{eq:lagrangian-first-principles}
\end{equation}
Deriving the presence or absence of separation from first principles is difficult: the velocity field $v$ itself should be a nonsmooth solution of the stochastic Navier--Stokes equations; many standard mathematical theories do not deal with generic (e.g. H\"older) non-Lipschitz vector fields $v$ (for Lipschitz velocity fields, there is no separation). Moreover, experimental verification is challenging, as it requires the tracking of particle pairs at very small distances, over a very broad range of temporal and spatial scale. Those difficulties have motivated the study of simplified models with suitably prescribed random nonsmooth velocity fields, often referred to as synthetic turbulence, both in physics and in mathematics.

One of the best-known examples is the nonsmooth Kraichnan model, introduced in \cite{kraichnan1968small}. In this model, the advecting velocity field is Gaussian, divergence-free, isotropic and nonsmooth in space, and white in time; its two-point motion can be explicitly computed, leading to particle separation and generalized stochastic flows, see e.g. \cite{bernard1998slow,vanden2000generalized, le2002integration}. Delta-correlation in time in the Kraichnan model is a simplifying choice, however unphysical, as in nature turbulent velocities are correlated in time; for this reason, more realistic models with time-correlated velocity fields have also been considered. 

However, when temporal correlations are retained, the situation is less understood. A popular example is the Ornstein--Uhlenbeck model with covariance given in Fourier by
\begin{equation}
    \widehat C(t-s,k)
    \simeq e^{-|k|^{2\beta}|t-s|}\,|k|^{-d-2\alpha},
    \label{eq:ou-covariance-introduction}
\end{equation}
where $\alpha$ controls the spatial roughness and $\beta$ the scale-dependent correlation time, see e.g. \cite{chaves2003lagrangian}. 
The numerical study of pair dispersion in this model and in other related synthetic velocity fields has been the subject of several studies in the past \cite{chaves2003lagrangian, thomson2005particle, devenish2009comment, eyink2013suppression, fung1992kinematic, fung1998two, nicolleau2011presence, sabelfeld1990numerical, elliott1995new}. Previous investigations, mostly conducted in the three-dimensional setting, have reported different scaling regimes, not always in agreement with Richardson's law, and with sensitivity to sweeping effects. Many of such works use a sparse approximation of \eqref{eq:ou-covariance-introduction} in Fourier space, often supported on dyadic frequencies, see e.g. \cite{sabelfeld1990numerical,fung1998two}, in order to control the computational cost.

In the two-dimensional case, the situation is less investigated and even less clear. Elliott and Majda \cite{elliott1996pair} consider an autonomous Gaussian velocity field (of the form \eqref{eq:ou-covariance-introduction} with $D_3=0$ and with an additional spatial mean), but supported along a fixed number of Fourier directions, reporting Richardson-like dispersion at large scales. Other numerical investigations \cite{fung1998two, nicolleau2003turbulent} for models like the Ornstein-Uhlenbeck one \eqref{eq:ou-covariance-introduction} also observe power-law particle dispersion; however, these results do not settle whether particle separation persists when the ultraviolet cutoff is removed before sending the initial particle distance to zero.

On the theoretical side, earlier works on Gaussian Markovian advection investigate the one-particle motion, see e.g. \cite{komorowski2004transport,fannjiang2002lagrangian}. Among them, in \cite{komorowski2004transport}, uniqueness for the one-point motion (that is, for the law of one particle) in an Ornstein--Uhlenbeck setting under the condition $\alpha+2\beta>1$ is proven.
Concerning pair separation, in \cite{bagnara2026regularity} the absence of Richardson dispersion is proven in the case of random autonomous continuous divergence-free fields in two dimensions, under a mild nondegeneracy assumption. Moreover, in the breakthrough paper \cite{colombo2026effective} almost-sure Lagrangian uniqueness for time-refreshing models is obtained when $\alpha+\beta>1$ (among a class of more general models); \cite{colombo2026effective} also gives a specific counterexample of a random velocity field exhibiting separation. However, to our knowledge, no analytical result is at present available for the Ornstein-Uhlenbeck model \eqref{eq:ou-covariance-introduction} outside the range $\alpha+\beta>1$. In a complementary direction, in the very recent result \cite{jimenez2026eddies} another model of random nonsmooth divergence-free velocity field is numerically investigated. For that model, \cite{fefferman2026richardson} gives a rigorous proof that the corresponding family of solutions shows both Richardson dispersion and dissipative anomaly is provided (see also \cite{de2026obukhov} for anomalous regularization).

In this work, we numerically investigate the problem of particle separation in a two-dimensional model with Gaussian velocity field $v$ of Ornstein-Uhlenbeck-type. The exact construction of $v$ is strongly inspired by the works by Elliott and Majda \cite{elliott1995new,elliott1996pair}, and it has similar regularity features as the field in \eqref{eq:ou-covariance-introduction}. We perform Lagrangian simulations and compute the mean-square particle distance. Our findings support mean-square particle separation for a specific choice of parameters in the complementary regime of \cite{colombo2026effective}: the extrapolated statistics are compatible with persistent separation at later times for the parameters studied; see Subsection \ref{subsec:distance_results} for more details.

The paper is organized as follows: Section~\ref{sec:velocity-field} defines the velocity ensemble; Section~\ref{sec:lagrangian-numerics} describes the numerical scheme and ordered extrapolations. Section~\ref{sec:numerical-results} presents the main numerical results, followed by brief conclusions. Appendices~\ref{app:elliotmajda}--\ref{app:torus-size} discuss covariance comparisons, resolution scales and rescaling.

\section{The stochastic velocity field}
\label{sec:velocity-field}

\subsection{The original Ornstein--Uhlenbeck model}

We first recall the original Ornstein--Uhlenbeck model for the velocity field as proposed in \cite{chaves2003lagrangian}. The velocity field
\[
    v:[0,T]\times\mathbb R^d\times\Omega\longrightarrow\mathbb R^d
\]
is a centered, divergence-free Gaussian field with covariance function
\[
    \mathbb E\bigl[v(t,x)v(s,y)^{\mathsf T}\bigr]
    =C(t-s,x-y).
\]
Writing the Fourier transform in the space variable as \(\widehat C_{\text{orig}}\), the covariance is given by
\begin{equation}
    \widehat C_{\text{orig}}(t-s,k)
    =D_2 e^{-D_3\langle k\rangle^{2\beta}|t-s|}
      \langle k\rangle^{-d-2\alpha}P_k^\perp,
    \qquad
    \langle k\rangle=\bigl(|k|^2+\ell_0^{-2}\bigr)^{1/2},
    \label{eq:ou-covariance}
\end{equation}
where \(\alpha>0\) and \(\beta>0\), and \(P_k^\perp\) is the Leray projection. 
In the following, we take $\ell_0^{-1}=1$ for simplicity.
Equivalently, $v$ is the stationary solution to the Ornstein--Uhlenbeck equation
\begin{equation}
    dv
    =-D_3(I-\Delta)^\beta v\,dt
     +\sqrt{2D_2D_3(I-\Delta)^{-d/2-\alpha+\beta}}\,d\widetilde W,
    \label{eq:ou-spde}
\end{equation}
where \(d\widetilde W\) is a divergence-free space--time white noise. In particular, the field $v$ satisfies, for every $\epsilon>0$,
\begin{equation}
    v\in C_tC_x^{\alpha-\epsilon}
       \cap C_t^{1/2-}C_x^{\alpha-\beta-\epsilon},\quad \PP\text{-a.s.}.
    \label{eq:regularity-statements}
\end{equation}
The choice \(D_3=0\) corresponds to the autonomous case (i.e. $v$ is time-independent). The Kraichnan regime, where $v$ is white in time, is recovered in the limit
\[
    D_2,D_3\longrightarrow\infty,
    \qquad
    D_1:=\frac{2D_2}{D_3}=\mathrm{const}.
\]
Indeed in this case we get
\begin{equation}
    \widehat C_{\text{orig}}(t-s,k)
    \longrightarrow
    D_1\delta_0(t-s)\langle k\rangle^{-d-2(\alpha+\beta)}P_k^\perp,
    \label{eq:kraichnan-limit}
\end{equation}
which is the covariance of a Kraichnan velocity field with exponent \(2(\alpha+\beta)\).

\subsection{Periodic formulation on the torus}\label{subsec:OU_torus}

The original Ornstein--Uhlenbeck model has two limitations in the present context: one needs to work with finitely many modes, and simulating all modes up to wavenumber \(N\) is computationally very expensive, with a cost \(O(N^2)\) per velocity evaluation with $d=2$. In this paragraph we address the first issue, introducing (in a very classical way) a generic Ornstein-Uhlenbeck model with independent modes on the torus of size $2L$.

We use the convention
\begin{align}
    &\mathbb T^d=[-L,L]^d,
    \qquad
    \widehat{\mathbb T}^d
    =\left(\frac{\pi}{L}\mathbb Z\right)^d,\\
    &\widehat f(k)=\int_{\mathbb T^d}f(x)e^{-ik\cdot x}\,dx,
    \qquad k\in\widehat{\mathbb T}^d,
    \label{eq:fourier-coeff}\\
    & f(x)
    =\frac{1}{(2\pi)^d}\left(\frac{\pi}{L}\right)^d
      \sum_{k\in\widehat{\mathbb T}^d}\widehat f(k)e^{ik\cdot x}
    =\frac{1}{(2L)^d}
      \sum_{k\in\widehat{\mathbb T}^d}\widehat f(k)e^{ik\cdot x}.
    \label{eq:fourier-inversion}
\end{align}

We consider a centered divergence-free Gaussian field
\[
    v:[0,T]\times\mathbb T^d\times\Omega\longrightarrow\mathbb R^d
\]
with covariance
\[
    \mathbb E\bigl[v(t,x)v(s,y)^{\mathsf T}\bigr]
    =C(t-s,x-y)
\]
whose Fourier coefficient
\begin{equation}
    \widehat C(t-s,k)
    =e^{-\mu(k)|t-s|}
      \frac{\widetilde\sigma(k)^2}{2\mu(k)}P_k^\perp,
    \qquad
    k\in\widehat{\mathbb T}^d\setminus\{0\}.
    \label{eq:torus-covariance}
\end{equation}
We assume in the following that $v$ has zero spatial average, that is
\begin{equation}
    \widehat{C}(0)=0.
\end{equation}
From the Fourier expansion of $C$, we get the following explicit representation for $v$:
\begin{equation}
    v(t,x)
    =2\sum_{k\in\widehat{\mathbb T}_+^d}
      \Bigl(a_k(t)\cos(k\cdot x)-b_k(t)\sin(k\cdot x)\Bigr),
    \label{eq:velocity-real-fourier}
\end{equation}
where $\widehat{\mathbb T}_+^d$ is the positive half-lattice
\[
    \widehat{\mathbb T}_+^d
    =\bigl\{k\in\widehat{\mathbb T}^d:
      k_d>0\ \text{or}\ k_d=0,\ k_{d-1}>0\ \text{or}\ \ldots k_d=k_{d-1}=\ldots k_2=0,\ k_1>0\}.
\]
and the processes \(a_k\)'s and \(b_k\)'s are \(\mathbb R^d\)-valued centered Gaussian processes with covariance function
\begin{equation}
    \mathbb E\bigl[a_k(t)a_k(s)^{\mathsf T}\bigr] = \mathbb E\bigl[b_k(t)b_k(s)^{\mathsf T}\bigr] = \frac{1}{2(2L)^d}\widehat C(t-s,k).
    \label{eq:fourier-covariance-ak}
\end{equation}
Equivalently, \(a_k\)'s and \(b_k\)'s are $\R^d$-valued independent stationary Ornstein-Uhlenbeck processes solution to
\begin{align}
    da_k
    &= -\mu(k)  a_k\,dt
     +\sigma(k)P_k^\perp\,d W^{k,a},\\
    db_k
    &= -\mu(k) b_k\,dt
     +\sigma(k)P_k^\perp\,d W^{k,b}
\end{align}
where $\sigma(k)=\frac{\widetilde\sigma(k)}{\sqrt{2}\,(2L)^{d/2}}$ and the $W^{k,a}$'s, and $W^{h,b}$'s are independent real Brownian motions. In particular, the $a_k$'s admit the following representation, at times \(t_j=j\Delta t\):
\begin{equation}
    a_k(0)\sim
    \mathcal N\!\left(0,
      \frac{\sigma(k)^2}{2\mu(k)}P_k^\perp
    \right),
    \label{eq:ak-initial}
\end{equation}
and
\begin{equation}
\begin{split}
    a_k(t_{j+1})
    ={}&e^{-\mu(k)\Delta t}a_k(t_j)\\
    &+\bigl(1-e^{-2\mu(k)\Delta t}\bigr)^{1/2}
      \frac{\sigma(k)}{\sqrt{2}\,\mu(k)^{1/2}}
      P_k^\perp\xi_{k,j,a},
\end{split}
    \label{eq:ak-update}
\end{equation}
where \(\xi_{k,j,a}\sim\mathcal N(0,I)\) are i.i.d. standard \(\mathbb R^d\)-valued Gaussian variables. The analogous representation holds for the $b_k$'s.

For \(d=2\), we can reduce \eqref{eq:ak-initial} and \eqref{eq:ak-update} to a scalar-valued representation:
\begin{equation}
    a_k(t)=z_{k,a}(t)\frac{k^\perp}{|k|},
    \qquad
    k^\perp=(-k_2,k_1),
    \label{eq:ak-zk}
\end{equation}
where $z_{k,a}$ satisfies
\begin{equation}
    z_{k,a}(0)\sim\mathcal N\!\left(0,\frac{\sigma(k)^2}{2\mu(k)}\right),
    \label{eq:zk-initial}
\end{equation}
and
\begin{equation}
    z_{k,a}(t_{j+1})
    =e^{-\mu(k)\Delta t}z_{k,a}(t_j)
     +\bigl(1-e^{-2\mu(k)\Delta t}\bigr)^{1/2}
      \frac{\sigma(k)}{\sqrt{2}\,\mu(k)^{1/2}}\xi_{k,j,a},
    \label{eq:zk-update}
\end{equation}
where $\xi_{k,j,a}$ are i.i.d. standard Gaussian variables. The analogous representation holds for the $(b_k)$'s.

In the autonomous case
\begin{equation}
    \widehat C(k)=\widetilde\sigma_A(k)^2P_k^\perp,
    \qquad
    k\in\widehat{\mathbb T}^d\setminus\{0\}.
    \label{eq:autonomous-covariance}
\end{equation}
a similar representation holds, with $\R^d$-valued independent variables $(a_k)$'s, $(b_k)'s$
\begin{equation}
    a_k,\,b_k\sim\mathcal N\bigl(0, \frac{\widetilde\sigma_A(k)^2P_k^\perp}{2(2L)^d}  \bigr),
    \label{eq:autonomous-ak}
\end{equation}
and, in dimension two,
\begin{equation}
    a_k=z_{k,a}
    \frac{k^\perp}{|k|},
    \qquad
    z_{k,a}\sim\mathcal N\bigl(0, \frac{\widetilde\sigma_A(k)^2}{2(2L)^2} \bigr),
    \label{eq:autonomous-zk}
\end{equation}
and analogously for the $b_k$'s.

\subsection{The sparse-mode model in dimension two}\label{subsec:ElliottMajda}

In this section, we introduce the reduced model for the velocity field, which we use in our simulations. The choice of this field is inspired by the construction in \cite{elliott1995new}, as well as by the sparse dyadic approximation in e.g. \cite{fung1998two,sabelfeld1990numerical}. Our aim is to retain both the spatial regularity and the temporal decorrelation of the original model in \eqref{eq:ou-covariance}, but substantially reduce the number of active modes. 
To this end, we retain only a subset of modes at each dyadic scale, selected along radial (approximate) directions. We rescale their amplitude so that the resulting field maintains comparable \(H^s\) and \(W^{s,p}\) norms with the model in \eqref{eq:ou-covariance}.

We work on the positive half-lattice equipped with the $\ell^\infty$ norm,
\begin{equation}
    \mathbb Z_+^2
    :=\bigl\{n\in\mathbb Z^2:\ n_2>0\ \text{or}\ n_2=0,\ n_1>0\bigr\},
    \quad
    \|n\|_\infty:=\max\{|n_1|,|n_2|\}.
    \label{eq:positive-lattice}
\end{equation}
We first introduce the dyadic squares
\begin{equation}
    \widetilde{\Delta}_j
    :=\frac{\pi}{L}\bigl( [-2^j,2^j]^2\cap\mathbb Z_+^2\bigr)
    \label{eq:deltasquare}
\end{equation}
and the dyadic blocks (or annuli)
\begin{equation}
    \Delta_1 = \widetilde{\Delta}_1,
    \qquad \Delta_j
    :=\widetilde{\Delta}_j \setminus \widetilde{\Delta}_{j-1} \quad \text{for }j\ge 2.
    \label{eq:deltaj}
\end{equation}
In particular, the first block contains modes with $1\le\|n\|_\infty\le2$, while the other blocks contain modes with $2^{j-1}<\|n\|_\infty\le2^j$.
The cardinality of a dyadic block satisfies
\begin{equation}
    \#\Delta_j
    =\frac32\,2^{2j}+2^j
    \approx 2^{2j}
    \label{eq:delta-cardinality}
\end{equation}
for $j\geq2$.
To each dyadic block, we associate its middle line
\begin{equation}
    Q_j
    :=\frac{\pi}{L}
      \left\{n\in \mathbb{Z}^2_+:\|n\|_\infty=m_j\right\},
    \label{eq:Qj}
\end{equation}
where $m_1=1$ and $m_j=(2^j+2^{j-1})/2$ for $j\ge 2$. Note that each $Q_j$ contains exactly $4m_j$ points.

Now we introduce the approximate directions where our modes are activated. For a fixed \(M\in\mathbb N_+\), the number of independent modes retained on each annulus $\Delta_j$ is $M_j=\min\{M,4m_j\}$ ($M$ morally represents the number of directions, but of course when $j$ is small we cannot take more than $4m_j$ points in each line $Q_j$).
We select distinct modes
\begin{equation}\label{eq:selected_modes}
    K_j=\bigl\{k^{j,0},k^{j,1},\ldots,k^{j,M_j-1} \bigr\}
\end{equation}
on \(Q_j\), approximately equispaced with respect to the discrete arclength along the positive half-perimeter of the square.
For $M=24$, the value used for our numerical investigation, we obtain $M_1=4$, $M_2=12$ and $M_j=24$ for $j\ge3$.
The modes are selected as follows: $Q_j$ is ordered in counterclockwise sense, starting from $k^{j,0}=(m_j,0)$, and, with this order, $k^{j,\ell}$ occupies position $\lceil 4m_jl/M_j\rceil$.

Following Elliott and Majda \cite{elliott1995new,elliott1996pair}, the covariance $C$ of our velocity field is supported only on the selected modes:
\begin{equation}
    \widehat C(t-s,k)=0
    \qquad\text{for }k\neq\pm k^{j,h},
    \label{eq:em-zero}
\end{equation}
On these modes, we impose
\begin{equation}
    \widehat C(t-s,k)
    =c(t-s,2^j\pi/L)P_k^\perp
    \qquad\text{for }k=\pm k^{j,h},
    \label{eq:em-covariance}
\end{equation}
with
\begin{align}
    c(t-s,2^j\pi/L)
    &=\frac{2^{2j}}{M}D_2
      e^{-D_3 (2^j\pi/L)^{2\beta}|t-s|}
      (2^j\pi/L)^{-(2+2\alpha)}
      \notag\\
    &=\frac{D_2}{M} (\pi/L)^{-2-2\alpha}
      e^{-D_3 (\pi/L)^{2\beta} 2^{2\beta j}|t-s|}
      2^{-2j\alpha}.
    \label{eq:cem}
\end{align}
In this way, we concentrate the spectral energy of the original model \eqref{eq:ou-covariance} in the activated wavenumbers $k_{j,h}$ of order $2^j$: at $t=s$ we have
\begin{equation}
    |\widehat C(0,k_{j,m})| \approx \#\Delta_j \cdot |\widehat C_{\text{orig}}(0,k)|\quad \text{for }k\in \Delta_j,
\end{equation}
and for general times $s,t$,
\begin{equation}
    |\widehat C(0,k_{j,m}) -\widehat C(t-s,k_{j,m})| \approx \#\Delta_j \cdot |\widehat C_{\text{orig}}(0,k) -\widehat C_{\text{orig}}(t-s,k)|\quad \text{for }k\in \Delta_j
\end{equation}
This makes the Gaussian velocity field with covariance $C$ comparable to the original one with covariance $C_{\text{orig}}$ in \eqref{eq:ou-covariance}. For more details, see Appendix~\ref{app:elliotmajda}, where we show that the two fields have similar regularity.

The model in \eqref{eq:em-zero},\eqref{eq:em-covariance},\eqref{eq:cem} is an Ornstein--Uhlenbeck model on the torus $\mathbb{T}^2$, as defined in Subsection \ref{subsec:OU_torus}, with drift and diffusion coefficients
\begin{equation}
    \mu^j(k)=D_3 (2^j\pi/L)^{2\beta},
    \qquad
    \widetilde\sigma(k)^2
    =\frac{2D_2D_3}{M}(\pi/L)^{-2-2\alpha+2\beta}2^{2j(-\alpha+\beta)}.
    \label{eq:em-ou-parameters}
\end{equation}
In particular, we can use the representation in Subsection \ref{subsec:OU_torus} for the corresponding velocity field $v$.

We remark that, as in \cite{elliott1995new,elliott1996pair}, and unlike in other works in the literature (e.g. \cite{fung1998two, thomson2005particle}), the modes $k^{j,\ell}$ are fixed and chosen in a deterministic way, in order to preserve Gaussianity of the velocity field $v$.

\section{Lagrangian model and numerical scheme}
\label{sec:lagrangian-numerics}

\subsection{The Lagrangian model and the separation question}
\label{subsec:lagrangian-question}
In this section, we describe precisely the question we address. Given our velocity field $v$, the Lagrangian dynamics is written formally as
\begin{equation}
    \dot X(t)=v(t,X(t)).
    \label{eq:lagrangian-formal}
\end{equation}
For the numerical implementation of \eqref{eq:lagrangian-formal}, we introduce a regularization of the velocity field, \(v^\epsilon\), through a Fourier cutoff. The two particle dynamics for the regularized model becomes 
\begin{equation}
\begin{aligned}
    \dot X^{\epsilon}(t)
    &=v^\epsilon\bigl(t,X^{\epsilon}(t)\bigr),
    &\qquad X^{\epsilon}(0)&=x,\\
    \dot Y^{\epsilon}(t)
    &=v^\epsilon\bigl(t,Y^{\epsilon}(t)\bigr),
    &\qquad Y^{\epsilon}(0)&=y.
\end{aligned}
\label{eq:regularized-particles}
\end{equation}We are interested in the behavior of
\[
    \E\bigl|X^{\epsilon}(t)-Y^{\epsilon}(t)\bigr|^2
\]
as \(\epsilon\to0\) first and \(|x-y|\to0\) then. In particular, our goal is to study whether the quantity
\begin{equation}
    \limsup_{|x-y|\to0}
    \limsup_{\epsilon\to0}
    \E\bigl|X^\epsilon(t)-Y^\epsilon(t)\bigr|^2,
    \qquad t>0.
    \label{eq:deterministic-separation-limsup}
\end{equation}
is zero or strictly positive. Here the ODEs \eqref{eq:lagrangian-formal} and \eqref{eq:regularized-particles} are considered on the full plane $\mathbb{R}^2$, with periodic field $v^\eps$ (equivalently, positions $X^\eps$, $Y^\eps$ are unwrapped lifts in $\mathbb{R}^2$ of trajectories driven by the field $v^\eps$ on $\T^2$).

\subsection{Numerical setup}
\label{subsec:numerical-setup}

In this section, we take the Gaussian centered velocity field $v$ with covariance function given by \eqref{eq:em-zero}, \eqref{eq:em-covariance}, \eqref{eq:cem}.
We introduce the time grid
\[
    t_\ell=\ell\,\Delta t .
\]
For $N\in\mathbb{N}$, $\eps=1/N>0$, we take the vector field \(v^N=v^\epsilon\) obtained imposing a UV Fourier cutoff on the field $v$ at wavenumber $\pi N/L$, that is, we set to zero all modes $k$ with $\|k\|_\infty>\pi N/L$.
Thanks to the representation in Subsection \ref{subsec:ElliottMajda}, we can build $n$ i.i.d. samples of $v^\eps$ at the grid times.
Namely, for each $i=1,\ldots n$, we take
\begin{equation}
    v^{\epsilon,(i)}(t_\ell,x)
    = 2 \sum_{|k|_\infty\le \pi N/L}
    \left[
      z_{k,a}^{(i)}(t_\ell)\frac{k^\perp}{|k|}\cos(k\cdot x)
      -z_{k,b}^{(i)}(t_\ell)\frac{k^\perp}{|k|}\sin(k\cdot x)
    \right],
    \label{eq:truncated-velocity}
\end{equation}
where the independent samples $z_{k,a}^{(i)}$ and $z_{h,b}^{(j)}$ are constructed as follows. We set
\begin{equation}
    z_{k,a}=0 \quad \text{for }k\notin\cup_j K_j
\end{equation}
(recall that $K_j$ are the selected modes at annulus $j$, see \eqref{eq:selected_modes}). For each $j$, for each Fourier mode \(k\) in $Q_j$, the initial values $z_{k,a}^{(i)}(0)$ are sampled independently from the stationary distribution of $z_{k,a}$, namely
\begin{equation}
    z_{k,a}^{(1)}(0),\ldots,z_{k,a}^{(n)}(0)
    \stackrel{\mathrm{i.i.d.}}{\sim}
    \mathcal N\!\left(0,\frac{D_2}
    {2M(2L)^2}(\pi/L)
    ^{-2-2\alpha}2^{-2j\alpha}\right),
    \label{eq:zc-initial}
\end{equation}
The coefficients $z_{k,a}^{(i)}$ are then evolved according to the following update:
\begin{equation}
\begin{split}
    z_{k,a}^{(i)}(t_{\ell+1})
    ={}&e^{-D_3 (2^j\pi/L)^{2\beta}\Delta t}
       z_{k,a}^{(i)}(t_\ell)\\
    &+\left(1-e^{-2D_3 (2^j\pi/L)^{2\beta}\Delta t}\right)^{1/2}
      \sqrt{\frac{D_2}
      {2M(2L)^2}}\,
      (\pi/L)^{-1-\alpha}2^{-j\alpha}\,
      \xi_{k,\ell,a}^{(i)},
\end{split}
    \label{eq:zc-update}
\end{equation}
where the random variables \(\xi_{k,\ell,a}^{(i)}\) are i.i.d. \(\mathcal N(0,1)\). The same construction is used for \(z_{k,b}^{(i)}\). For $D_3=0$, the case of an autonomous velocity field, the same equal-time Gaussian field is sampled once and then fixed.

For each realization \(v^{\epsilon,(i)}\), we consider two particles \((X^{\epsilon,(i)})_\ell\) and \((Y^{\epsilon,(i)})_\ell\) starting from
\begin{equation}
    x=d_0e_1,
    \qquad
    y=0,
    \label{eq:numerical-initial-data}
\end{equation}
solving the ODE \eqref{eq:regularized-particles} with the same velocity realization $v^\epsilon=v^{\epsilon,(i)}$. The trajectories \((X^{\epsilon,(i)})_\ell\) and \((Y^{\epsilon,(i)})_\ell\) are approximated using an explicit two-stage predictor--corrector (Heun) scheme, with the same timestep $\Delta t$; we call $X_{t_\ell}^{\epsilon,\Delta t,(i)}$, $Y_{t_\ell}^{\epsilon,\Delta t,(i)}$ the corresponding simulated paths.

Standard Monte Carlo is then used to estimate the mean-square particle separation at a prescribed final time \(T\), namely
\begin{equation}
    \mathbb E\bigl[|X_T^\epsilon-Y_T^\epsilon|^2\bigr] \approx \frac{1}{n}\sum_{i=1}^n |X_T^{\epsilon,\Delta t,(i)}-Y_T^{\epsilon,\Delta t,(i)}|^2
    \label{eq:monte-carlo-target}
\end{equation}
Independent ensembles are generated for different final times, and the sample standard deviation is also estimated to quantify the variability of the simulated separations. 

Finally, we emphasize that the model \eqref{eq:em-zero}, \eqref{eq:em-covariance}, \eqref{eq:cem} is treated as the model of interest in its own right, rather than as an approximation of the original model \eqref{eq:ou-covariance}. For this reason, the number \(M\) of selected modes per dyadic scale is kept fixed.

\subsection{Order of limits and numerical extrapolation}
\label{subsec:order-limits}

Our analysis involves three successive limiting procedures, which are taken in the following order:
\[
    \Delta t\to0,
    \qquad
    \epsilon=\frac1N\to0,
    \qquad
    d_0=|x-y|\to0.
\]
Ideally, we would like to: a) for every fixed \(\epsilon\) and \(d_0\), choose \(\Delta t=\Delta t(\epsilon,d_0)\) sufficiently small so that the particle-separation statistics \(|X^\epsilon-Y^\epsilon|\) become stable; b) then, for every \(d_0\), choose \(\epsilon=\epsilon(d_0)\) sufficiently small so that the same quantity is stable; finally, c) choose \(d_0\) small enough so that again this quantity is stable.

Since reaching sufficiently small values of all parameters can be computationally demanding, we employ numerical extrapolation to estimate the corresponding limits. More precisely, to estimate numerically
\[
    \lim_{a\to0}f(a)
\]
from values \(f(a_j)\), \(j=1,\ldots,m\), we fit the data with a polynomial curve \(g\), by weighted least squares. Its intercept $g(0)$ is then taken as the estimate. We use a quadratic polynomial for the extrapolations in $\Delta t$ and $\epsilon= 1/N$; this choice for extrapolation in $\Delta t$ is consistent with the fact that the integrator is formally of second order; for extrapolation in $\epsilon$, this choice is an Ansatz. For the extrapolation in $d_0 \to 0$, we compare both linear and quadratic fits.

The use of numerical extrapolation also allows to get an uncertainty estimate for the extrapolated value. The statistical errorbars, for each data set with given $\Delta t$, $\epsilon$, $d_0$, consist in one bootstrap standard error, obtained by resampling the samples with replacement; the weighted fits employ those errors as input standard deviations to estimate the standard deviation of the fit $\Delta t\to 0$. The estimated standard deviations are then used as input to estimate the standard deviation of the fit $\epsilon \to 0$. This procedure is repeated once more to estimate the standard deviation of the fit $d_0 \to 0$.

%All errorbars

No blocking procedure is necessary as all datapoints corresponding to different parameters are independent, as they are obtained from different samples.

Before doing the extrapolation, we need to take into account some constraints on the scales, in order to avoid trivialities.
To adequately resolve the spatial and temporal scales of the regularized velocity field, the initial particle separation should satisfy
\begin{equation}
    \ell_\epsilon\ll d_0,
    \label{eq:lipschitz-vs-separation}
\end{equation}
where $\ell_\epsilon$ denotes the UV cutoff, in order to place \(|x-y|\) in the non-smooth spatial scale, namely the particle separation probes spatial scales beyond the regularization-induced smooth regime. In addition, the timestep is expected to satisfy
\begin{equation}
    \Delta t\ll
    \min\{\delta t_{\mathrm{strain}},\delta t_{\mathrm{sweep}},\delta t_{\mathrm{corr}}\},
    \label{eq:timestep-scales}
\end{equation}
where \(t_{\mathrm{strain}}\), \(t_{\mathrm{sweep}}\), \(t_{\mathrm{corr}}\) denote respectively the inverse-strain timescale, the sweeping timescale and the correlation timescale. In Appendix~\ref{app:scales} we provide reasonable estimates for $\ell_\eps$, $V$ and $\delta t_\eps$, based on the scaling properties of the original Ornstein-Uhlenbeck model \eqref{eq:ou-covariance}. The conditions \eqref{eq:lipschitz-vs-separation} and \eqref{eq:timestep-scales} become
\begin{equation}
    L\eps \ll d_0,
    \label{eq:expected-lipschitz-scale}
\end{equation}
and
\begin{equation}
    \Delta t
    \ll
    \min\left\{
      \frac{L^{1-\alpha}}{\sqrt{D_2}}\eps,\,
      \frac{L^{2\beta}}{D_3}\eps^{2\beta}
    \right\}.
    \label{eq:expected-timestep-scale}
\end{equation}
In the numerical implementation, we retain the factors associated to the physical wavenumbers and the selected modes, while the computation presented here describes only the scaling of the quantities, up to multiplicative constants.

We also monitor a CFL-type advective resolution condition at each time step,
\[
  C_n = \Delta t\,k_{\max}
  \max_{\substack{i=1,2\\s=1,2}}
  \left|\mathbf{v}^{(s)}_i\right|
  \le C_{\max}, \qquad C_{\max}=0.5,
\]
where $k_{\max}$ is the largest modulus of the retained physical Fourier wavevectors, and $s$ labels the two stages of the Heun method. 
This condition is combined with the condition \eqref{eq:expected-timestep-scale} above.

\section{Numerical results}\label{sec:numerical-results}

In this section we report the results of our main investigation: the study of tracers advected by the time-dependent stochastic velocity field $v$ with covariance defined in \eqref{eq:em-zero},\eqref{eq:em-covariance},\eqref{eq:cem}.

As in the section above, we mainly focus on the mean square particle distance, $\langle r^2 \rangle$, but we also compute the probability density function of the distance $r = |X-Y|$ at different times.

\subsection{Extrapolation results}

We first investigate the dependence of the numerical results on the discretization parameters, following the order of limits discussed in Section~\ref{subsec:order-limits}.

\begin{figure}[htbp]
    \centering

    % (a) Timestep extrapolation
    \begin{subfigure}[t]{0.49\textwidth}
        \centering
        \includegraphics[width=\linewidth]{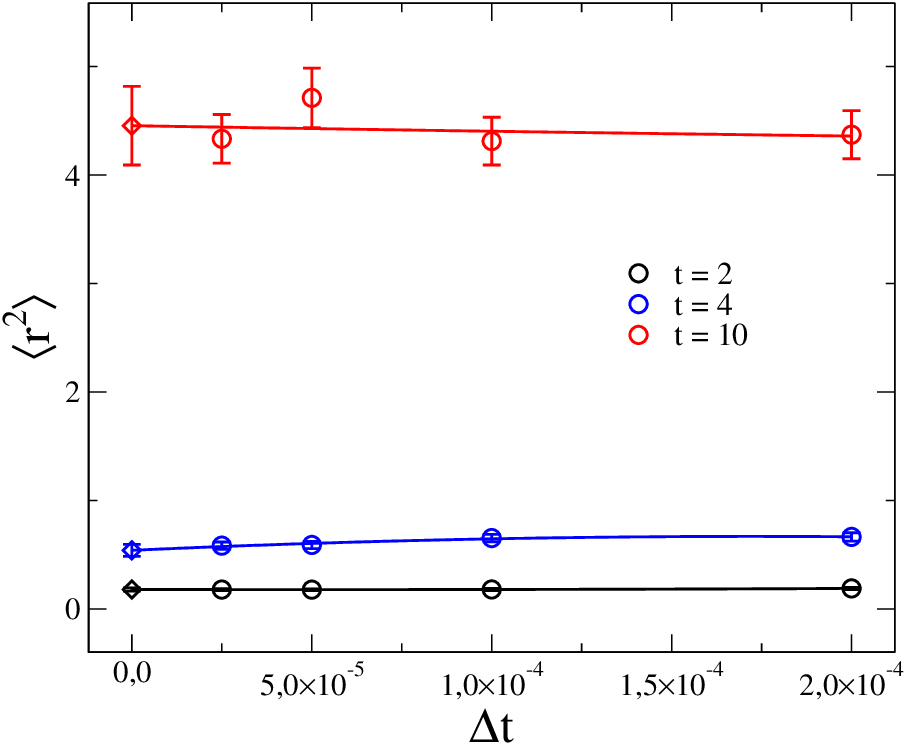}
        \caption{}
        \label{fig:dt-extrapolation}
    \end{subfigure}
    \hfill
    % (b) Ultraviolet cutoff extrapolation
    \begin{subfigure}[t]{0.49\textwidth}
        \centering
        \includegraphics[width=\linewidth]{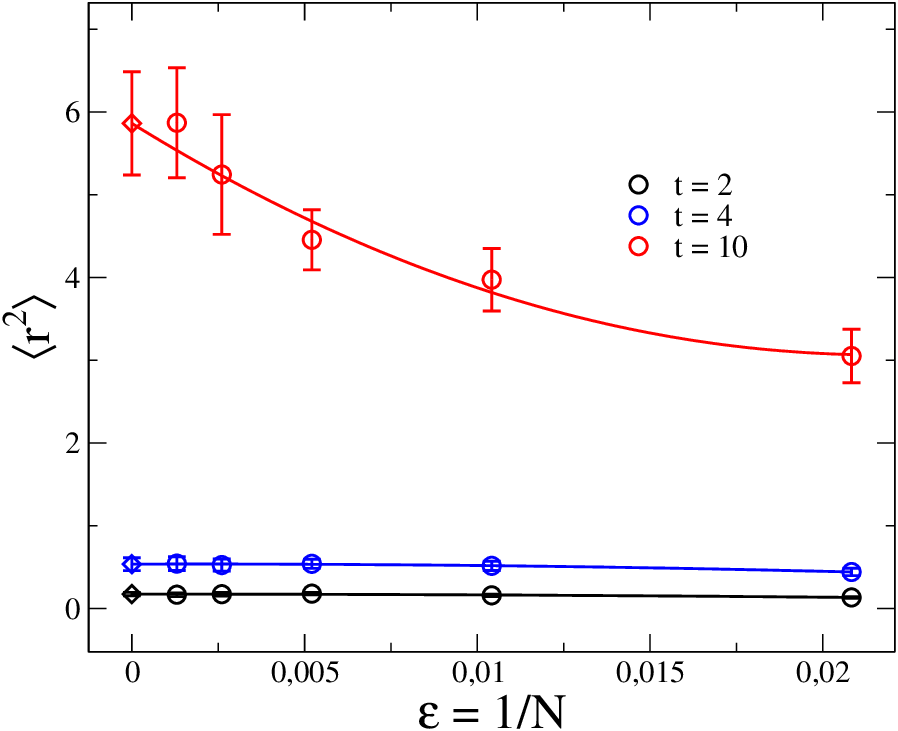}
        \caption{}
        \label{fig:epsilon-extrapolation}
    \end{subfigure}

    \vspace{0.5cm}

    % (c) Initial separation extrapolation
    \begin{subfigure}[t]{0.49\textwidth}
        \centering
        \includegraphics[width=\linewidth]{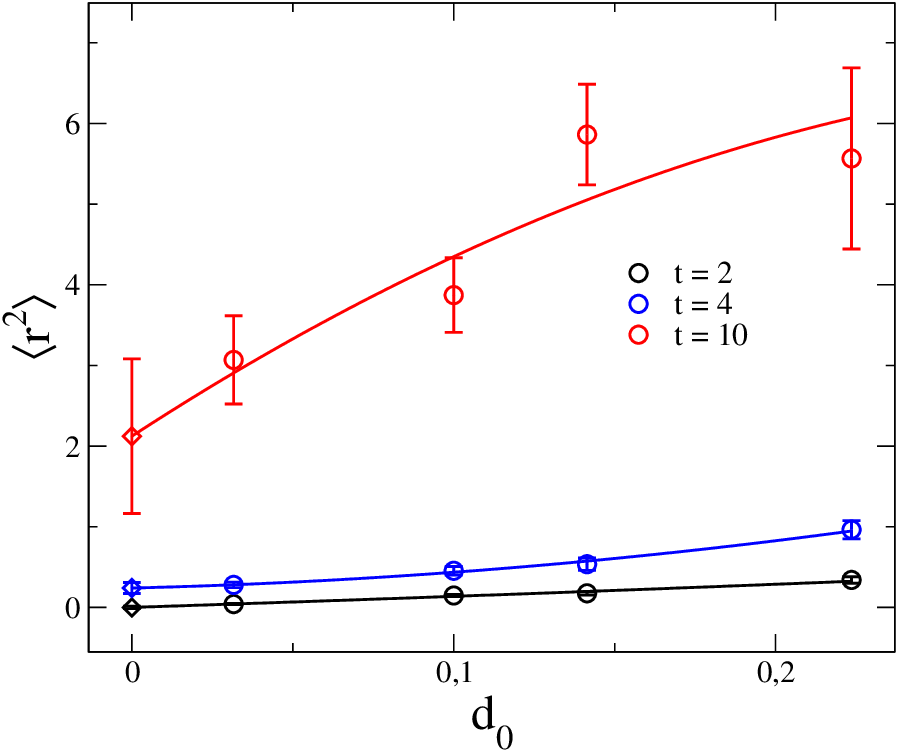}
        \caption{}
        \label{fig:d0-extrapolation}
    \end{subfigure}

    \caption{
    \textbf{(a)} Mean-square particle distance as a
    function of the timestep $\Delta t$, for $N=192$
    and $d_0^2=0.02$, at three different final times.
    The intercepts of the fits represent the
    extrapolated values (indicated with diamonds).\\
    \textbf{(b)} Mean-square particle distance as a
    function of the inverse ultraviolet cutoff
    $\epsilon$, for $d_0^2=0.02$, at three different
    final times. The intercepts of the fits represent
    the extrapolated values (indicated with diamonds).\\
    \textbf{(c)} Mean-square particle distance as a
    function of the initial distance $d_0$, at three
    different final times. The intercepts of the fits
    represent the extrapolated values (indicated with
    diamonds).
    }
    \label{fig:extrapolations}
\end{figure}

We start by studying the convergence with respect to the timestep $\Delta t$. In Figure~\ref{fig:dt-extrapolation}, we report the mean-square distance at different times for a fixed ultraviolet cutoff $N=192$ and fixed square initial distance $d_0^2=0.02$ for different final times.
The plot shows a relatively weak dependence on the timestep over the investigated range, with variations comparable to the statistical uncertainties. The extrapolated values at $\Delta t=0$ are obtained from a quadratic fit of the data, as the integrator used is, formally, of second order. However, the fits are consistent with a linear behavior in $\Delta t$; this may be expected as after the evolution of the Fourier coefficients the second order may not be maintained.

We next consider the dependence on the ultraviolet cutoff, and we extrapolate the mean-square distance towards $\epsilon=0$, corresponding to the removal of the Fourier cutoff.
Figure~\ref{fig:epsilon-extrapolation} shows the mean-square distance as a function of the inverse cutoff parameter $\epsilon$, again for different final times and square initial distance $d_0^2= 0.02$.

Finally, we investigate the dependence on the initial distance $d_0$. We report our results in Figure~\ref{fig:d0-extrapolation}: the mean-square distance is plotted for different initial distances. From Figure~\ref{fig:d0-extrapolation}, we see that the fitted intercepts at $d_0=0$ are positive at the larger times: this is compatible with the possibility of a nontrivial separation in the limit $d_0 \to 0$. The extrapolated limits for $d_0\to 0$, for other final times, are presented and discussed in more detail in the next Subsection.

\subsection{Mean-square particle distance}\label{subsec:distance_results}

\begin{figure}[htbp]
    \centering
    \includegraphics[width=0.8\linewidth]{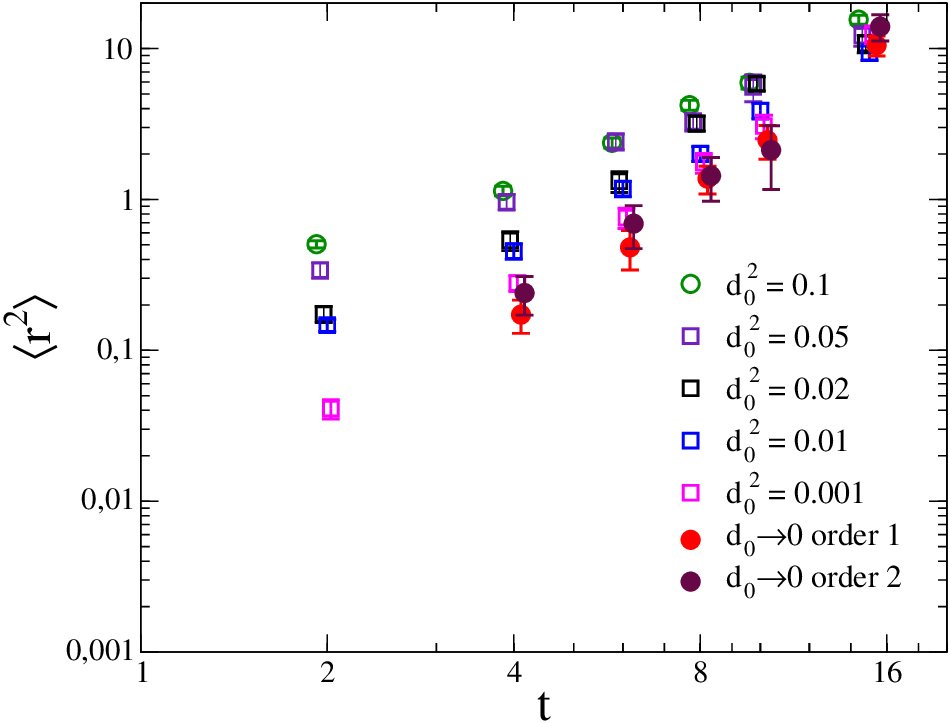}
    \caption{Results for the two-point motion of tracer particles in a time-correlated stochastic velocity field, $\alpha = 0.2$, $\beta = 0.5$, $D_3 = 1$, $D_2 = 5$, $L = 20$, $M = 24$. 
    Order one and two refer to the order of the polynomial used for the fit after timestep and cutoff extrapolation (data have been slightly shifted on the horizontal axis to improve the readability). }
    \label{fig:beta-extrapolated}
\end{figure}

In Figure~\ref{fig:beta-extrapolated} we report our extrapolated results for the time-correlated model, as the initial distance $d_0$ goes to $0$, for different final times. This represents the main goal of our work.

As we see from Figure~\ref{fig:beta-extrapolated}, starting from final time $t=4$, and up to $t=15$, all extrapolated limits as $d_0\to 0$, both with linear and quadratic fits, are compatible with
\begin{equation}
    \limsup_{|x-y|\to0}
    \limsup_{\epsilon\to0}
    \E\bigl|X^\epsilon(t)-Y^\epsilon(t)\bigr|^2 >0,
\end{equation}
that is, particle separation in mean-square.

The limit for final time $t=2$ instead is compatible with zero value (it is not shown in the figure, due to the log scale in the vertical axis). This could be related to a transient phase, where, for very small values of $d_0$, the two point distance is still too close to the initial distance $d_0$.

All simulations are run on the torus with size $L=20$. In Appendix \ref{app:torus-size}, we discuss the influence of the torus size $L$ and of the final time $T$ on the results.

\subsection{Probability density functions}

\begin{figure}[htbp]
    \centering

    \begin{subfigure}[t]{0.49\textwidth}
        \centering
        \includegraphics[width=\linewidth]{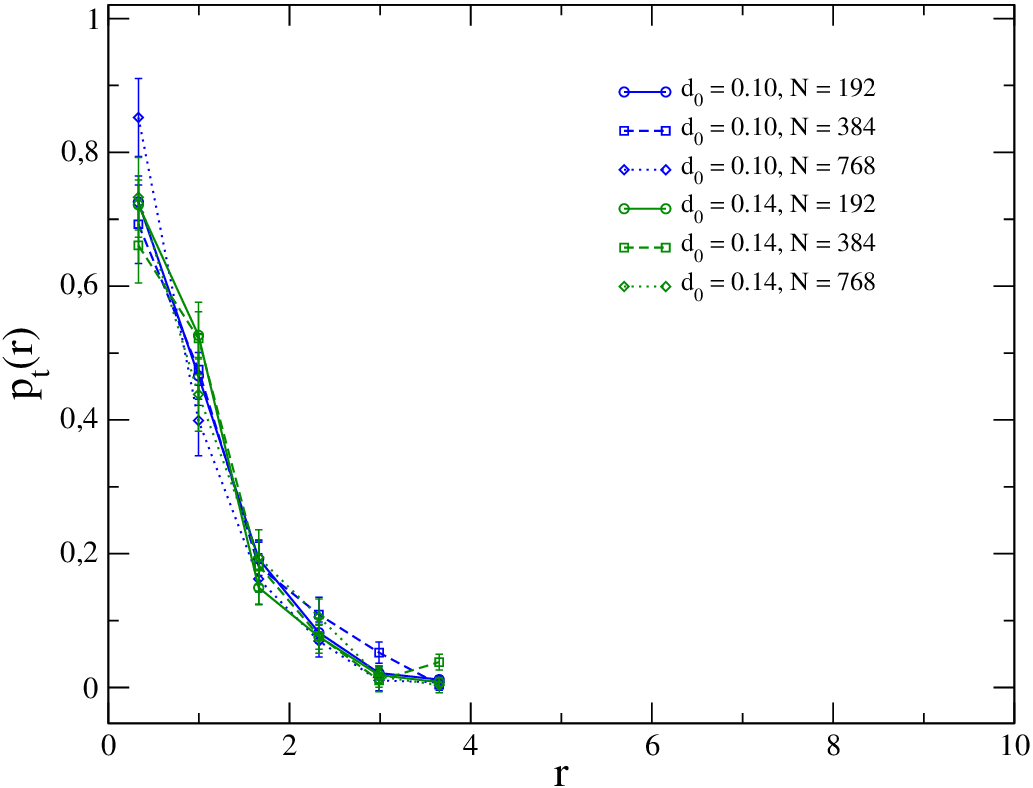}
        \caption{$t=6$.}
        \label{fig:pdf-t6}
    \end{subfigure}
    \hfill
    \begin{subfigure}[t]{0.49\textwidth}
        \centering
        \includegraphics[width=\linewidth]{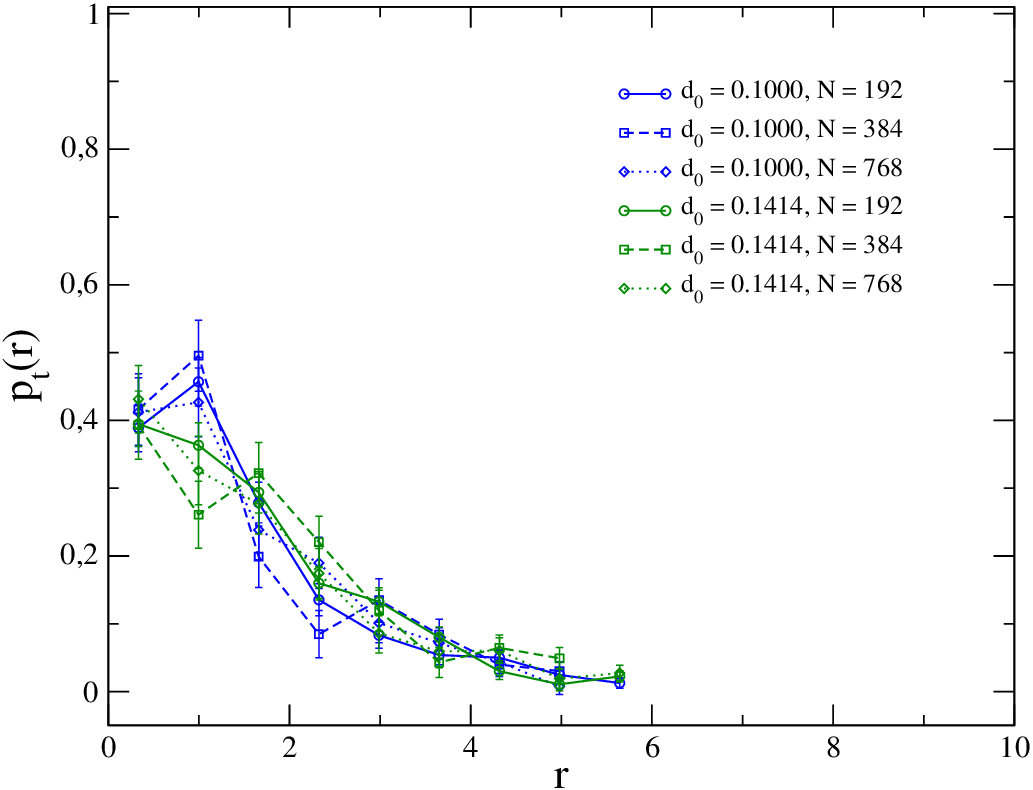}
        \caption{$t=10$.}
        \label{fig:pdf-t10}
    \end{subfigure}

    \caption{
    Probability densities of the particle separation after timestep
    extrapolation, for $d_0^2=0.01,0.02$ and $N=192,384,768$.
    Errorbars were obtained via bootstrap procedure.
    }
    \label{fig:pdf-comparison}
\end{figure}

We also investigate the (empirical) probability density function $p_t(r)$ of the particle distance $|X_t-Y_t|$ at time $t$. Values of $N$ and $d_0$ are fixed for each curve. 
Figures~\ref{fig:pdf-t6} and~\ref{fig:pdf-t10} show estimates obtained from histograms on common bins, extrapolated pointwise as $\Delta t\to0$, for different ultraviolet cutoffs and initial distances.

At $t=6$, the probability densities 
are stable in the UV cutoff and initial distance, and appear to be more concentrated in $r=0$. At $t=10$, they are instead more disperse. Some dependence on the cutoff and the initial distance remains visible, particularly near the peak, at $t=10$.

\section{Concluding Remarks}

In this work, we numerically investigated the particle pair dispersion in a two-dimensional time-correlated Gaussian velocity field, with sparse energy spectrum, using successive extrapolations in the timestep $\Delta t$, the UV cutoff $1/N$, and the initial condition $d_0$.

For this model, the extrapolated mean-square separations are compatible with particle separation at positive time, namely the limit of the mean-square separation stays positive as the UV cutoff is removed and the initial distance $d_0$ is sent to zero. These numerical findings suggest that particle separation may occur in this model, under the investigated range of parameters.

A natural direction for future work is to investigate whether the mean-square particle separation exhibits a Richardson-type power-law scaling in time and, if so, to determine the corresponding exponent.

% Paper-style appendices drafted only from the handwritten notes
% "Notes appendix(1).pdf".
% Requires at least: amsmath, amssymb.

 \appendix

\section{Covariance carried by the sparse model}
\label{app:elliotmajda}

In this appendix, we compare our covariance $C$ \eqref{eq:em-zero},\eqref{eq:em-covariance},\eqref{eq:cem} introduced in Subsection covariance in \eqref{eq:ou-covariance}, showing a similar regularity behaviour. The computations are all standard. For simplicity, we take $2L=1$ here.

To be more precise, as the original model is defined on the full space, we introduce the natural analogue model on the torus $\T^d$, namely we consider the power-law covariance
\begin{equation}
     \widehat C_{\mathrm{pl}}(t-s,k)
    =D_2e^{-D_3|k|^{2\beta}|t-s|}|k|^{-2-2\alpha}P_k^\perp
    =c_{\mathrm{pl}}(t-s,k)P_k^\perp,\quad k\in \widehat{\T}^d \backslash \{0\}.
    \label{eq:pl-covariance}
\end{equation}
We are going to compare the regularity of the velocity field $v$ of our model with that of the Gaussian velocity field $v_{\mathrm{pl}}$ with covariance $C_{\mathrm{pl}}$.

We recall that a random centered Gaussian velocity field $u$ with covariance $K$, defined on the torus (and for simplicity with zero spatial mean), admits the Fourier representation
\begin{equation}\label{eq:random_field_Fourier}
    u(x) = 2\sum_{k\in \widehat\T^d_+} z_k e^{ik\cdot x}
\end{equation}
with $z_k$, $k\in\mathbb{T}^d_+\cup\{0\}$ i.i.d. centered Gaussian variables with $\E[z_kz_k^\top]=\widehat K(k)/2$. The homogeneous Sobolev norm of $u$ satisfies
\begin{equation}
    \E\|u\|_{\dot{H}^s}^2 = 2\E \sum_{k\in \widehat\T^d_+} |k|^{2s}|z_k|^2 = \sum_k |k|^{2s} \operatorname{Tr} \widehat{K}(k).
\end{equation}
For our velocity field $f$, we note that
\begin{align}
    \sum_{k\in\Delta_j}\operatorname{Tr} \widehat C(0,k) = M_jc(0,2^j\pi/L)
    &\approx 2^{2j}D_2 2^{-j(2+2\alpha)}
    \notag\\
    &\approx \sum_{k\in\Delta_j}D_2|k|^{-(2+2\alpha)}
    =\sum_{k\in\Delta_j}\operatorname{Tr} \widehat C_{\mathrm{pl}}(0,k).
    \label{eq:covariance-shell-comparison}
\end{align}
and also that
\begin{align}
    \sum_{k\in\Delta_j}\operatorname{Tr}[\widehat C(0,k)-\widehat C(t-s,k)] &= M_j(c(0,2^j\pi/L)-c(t-s,2^j\pi/L)) \notag\\
    &\approx 2^{2j}D_2 (1-e^{-D_3 2^{2\beta j}|t-s|}) 2^{-j(2+2\alpha)}
    \notag\\
    &\approx \sum_{k\in\Delta_j}D_2 (1-e^{-D_3 |k|^{2\beta}|t-s|}) |k|^{-(2+2\alpha)} \notag\\
    &=\sum_{k\in\Delta_j}\operatorname{Tr}[\widehat C_{\mathrm{pl}}(0,k)-\widehat C_{\mathrm{pl}}(t-s,k)].
    \label{eq:covariance-shell-comparison-2}
\end{align}
In other words, the total spectral energy over an annulus is comparable in the two models. Therefore, we get comparable Sobolev regularity. For the homogeneous Sobolev norm, 
\begin{align}
    \mathbb E\|v(t)\|_{\dot H^s}^2
    &=\sum_k |k|^{2s} \operatorname{Tr}\widehat{C}(0,k)\\
    &\approx 2D_2\sum_j2^{2j(s-\alpha)}
      \notag\\
    &\approx 2\sum_j\sum_{k\in\Delta_j}
      |k|^{2s} \operatorname{Tr}\widehat{C}_{\mathrm{pl}}(0,k)
      =\mathbb E\|v_{\mathrm{pl}}(t)\|_{\dot H^s}^2.
    \label{eq:sobolev-comparison}
\end{align}
Likewise, for temporal increments,
\begin{align}
    \mathbb E\|v_{\mathrm{EM}}(t)-v_{\mathrm{EM}}(s)\|_{\dot H^s}^2
    &=4\sum_k|k|^{2s}
      \operatorname{Tr}\left(\widehat C(0,k) -\widehat C(t-s,k)\right)
      \notag\\
      &\approx 4\operatorname{Tr}\left(\widehat C_{\mathrm{pl}}(0,k) -\widehat C_{\mathrm{pl}}(t-s,k)\right)\\
    &=\mathbb E\|v_{\mathrm{pl}}(t)-v_{\mathrm{pl}}(s)\|_{\dot H^s}^2 \notag\\
    &\lesssim \sum_j2^{2j(s-\alpha+\beta)}|t-s|
    \label{eq:sobolev-increment-comparison}
\end{align}
A similar comparison can be carried out in Besov spaces.
For \(p\) an even integer, exploiting the representation \eqref{eq:random_field_Fourier} with time-dependent coefficients $z_k$ (for $v$) and $z_k^{\mathrm{pl}}$ (for $v_{\mathrm{pl}}$), using Gaussianity, stationarity and independence, we obtain
\begin{align}
    \mathbb E\|v(t)\|_{B_{p,p}^s}^p
    &=\,\mathbb E\sum_j2^{jsp}
      \left\|\sum_{k\in\Delta_j}
      z_k(0)e^{ik\cdot x}\right\|_{L^p}^p
      \notag\\
    &\overset{\mathrm{indep.}}{\approx_p}
      \sum_j2^{jsp}
      \sum_{k_1,\ldots,k_{p/2}\in\Delta_j}
      \mathbb E\left[
      z_{k_1}(0)^2\cdots
      z_{k_{p/2}}(0)^2\right]
      \notag\\
    &\overset{\mathrm{Gaussianity}}{\approx}
      \sum_j2^{jsp}M_j^{p/2}
      c(0,2^j)^{p/2}
      \notag\\
    &\approx \sum_j2^{jsp}
      \left(\sum_{k\in\Delta_j}c_{\mathrm{pl}}(0,k)\right)^{p/2}
      \notag\\
    &\overset{\mathrm{Gaussianity}}{\approx}
      \sum_j2^{jsp}
      \sum_{k_1,\ldots,k_{p/2}\in\Delta_j}
      \mathbb E\left[
      z_{k_1}^{\mathrm{pl}}(0)^2\cdots
      z_{k_{p/2}}^{\mathrm{pl}}(0)^2\right]
      \notag\\
    &\overset{\mathrm{indep.}}{=}\mathbb E\|v_{\mathrm{pl}}(t)\|_{B_{p,p}^s}^p.
    \label{eq:besov-comparison}
\end{align}
Similarly, for temporal increments, we obtain
\begin{align}
    \mathbb E\|v(t)-v(s)\|_{B_{p,p}^s}^p
    &=\,\mathbb E\sum_j2^{jsp}
      \left\|\sum_{k\in\Delta_j}
      \bigl(z_k(0)-z_k(t-s)\bigr)
      e^{ik\cdot x}\right\|_{L^p}^p
      \notag\\
    &\approx
      \mathbb E\|v_{\mathrm{pl}}(t)-v_{\mathrm{pl}}(s)\|_{B_{p,p}^s}^p.
    \label{eq:besov-increment-comparison}
\end{align}

\section{Spatial scales and time scales}
\label{app:scales}

This appendix section is devoted to the estimate of the characteristic spatial scale and the temporal scale of the regularized velocity field \(v^\epsilon\). We do not work directly with our model \eqref{eq:em-zero},\eqref{eq:em-covariance},\eqref{eq:cem}, but rather with the power-law model \eqref{eq:pl-covariance}, or more precisely with its UV truncated version at wavenumber $\pi \eps^{-1}/L$. The model \eqref{eq:pl-covariance} shares similar regularity features of our model, as we have seen, and has the advantage here of allowing explicit computations.

We first estimate the spatial scale. The spatial scale corresponding to the UV Fourier cutoff at \(\pi/(\eps L)\) is
\begin{equation}
    \ell_\eps \approx L\eps.
\end{equation}

Concerning the temporal scales, the inverse-strain scale is the timescale $\delta t_{\mathrm{strain}}$ such that
\begin{equation}
    \delta t_{\mathrm{strain}}|\nabla v^\epsilon|\approx 1.
\end{equation}
The mean-square gradient of $v^\epsilon$ has order
\begin{equation}
\begin{split}
    \E|\nabla v^\epsilon|^2 &= D^2C_{\mathrm{pl}}^\epsilon(0) \\
    &\approx
    \int_{1/L\le |k|\le 1/(L\epsilon)}
    D_2|k|^{-2\alpha}\,dk \\
    &\approx D_2(L\epsilon)^{-2+2\alpha}.
\end{split}
\end{equation}
We therefore have
\begin{equation}
    \delta t_{\mathrm{strain}} \approx \frac{(L\eps)^{1-\alpha}}{\sqrt{D_2}}.
\end{equation}
The sweeping scale $\delta t_{\mathrm{sweep}}$ is the typical time needed for a particle to reach the spatial UV cutoff $\ell_\epsilon$, namely
\begin{equation}
    \delta t_{\mathrm{sweep}}|v^\epsilon|\approx \ell_\epsilon.
\end{equation}
The mean-square $v^\epsilon$ has order
\begin{equation}
\begin{split}
    \E|v^\epsilon|^2 &= C_{\mathrm{pl}}^\epsilon(0) \\
    &\approx
    \int_{1/L\le |k|\le 1/(L\epsilon)}
    D_2|k|^{-2-2\alpha}\,dk \\
    &\approx D_2 L^{2\alpha}.
\end{split}
\end{equation}
Hence we get
\begin{equation}
    \delta t_{\mathrm{sweep}} \approx \frac{L\eps}{\sqrt{D_2}L^\alpha} = \frac{L^{1-\alpha}\eps}{\sqrt{D_2}}.
\end{equation}
Finally, the correlation scale $\delta t_{\mathrm{corr}}$ is the time needed for $v^\epsilon$ to decorrelate. We must then impose, for all active wavenumbers $k$,
\begin{equation}
    D_3|k|^{2\beta}\delta t_{\mathrm{corr}}\lesssim1.
    \label{eq:oscillation-condition}
\end{equation}
The most restrictive condition is obtained at the maximal wavenumber \( |k_{\max}|\approx(L\epsilon)^{-1} \), giving
\begin{equation}
    \delta t_{\mathrm{corr}}
    \approx
    \frac1{D_3|k_{\max}|^{2\beta}}
    \approx
    \frac1{D_3(1/(L\epsilon))^{2\beta}}
    =\frac{L^{2\beta}}{D_3}\epsilon^{2\beta}.
    \label{eq:oscillation-time-scale}
\end{equation}

\section{Influence of the torus size}
\label{app:torus-size}

One may ask whether the torus size $L=20$ and final times up to $T=15$ are appropriate choices to observe presence or absence of separation.
As the model is still poorly understood from a theoretical viewpoint, we cannot give a fully satisfactory answer to this question, but let us note two points. First, while in principle separation, if true, could be observed for smaller times, there must be a transient phase in simulations (depending on the smallest initial distance we can reach), where separation cannot be observed even if true (this could be the case for final time $T=2$). Second, as we will now show, $T=15$ seems a good candidate as typical time scale for $L=20$, in the following sense: when we rescale time and space to be of order one, the coefficients $D_2$ and $D_3$ of the scaled model are still of order one. This further motivates the choice of the current space- and time-scales.

To show this, we rescale the Lagrangian dynamics to a unit spatial domain and a unit time interval, and we derive the corresponding expressions for the scaled parameters \( \widetilde D_2\) and \( \widetilde D_3\). Finally, we verify that the parameter values used in our experiments correspond, after rescaling, to covariance coefficients \( \widetilde D_2\) and \( \widetilde D_3\) of order one. All computations are performed for the power-law covariance $C_{\mathrm{pl}}$ given in \eqref{eq:pl-covariance}, which shares similar regularity properties to our model \eqref{eq:em-zero},\eqref{eq:em-covariance},\eqref{eq:cem}.

We consider the Lagrangian dynamics of a tracer,
\begin{equation}
    \dot x=v(t,x).
    \label{eq:original-scaled-ode}
\end{equation}
on the torus \(\mathbb T_L^d\) of size \(L\) up to final time \(T\). We rescale space and time, in order to  normalize to one both the torus half-size and the final time. To this end, we introduce the rescaled quantities 
\begin{equation}
    \widetilde x(\widetilde t)
    =\frac1L x(T\widetilde t),
    \qquad
    \dot{\widetilde x}(\widetilde t)
    =\widetilde v\bigl(\widetilde t,\widetilde x(\widetilde t)\bigr).
    \label{eq:rescaled-trajectory}
\end{equation}
The rescaled velocity then satisfies
\begin{equation}
    \widetilde v\bigl(\widetilde t,\widetilde x(\widetilde t)\bigr)
    =\frac{T}{L}\dot x(T\widetilde t)
    =\frac{T}{L}
      v\bigl(T\widetilde t,L\widetilde x(\widetilde t)\bigr).
    \label{eq:rescaled-velocity-as-written}
\end{equation}
Consequently, the corresponding covariance is given by
\begin{equation}
\begin{split}
    \widetilde C_{\mathrm{pl}}(
      \widetilde t-\widetilde s,
      \widetilde x-\widetilde y)
    &=\mathbb E\left[
      \widetilde v(\widetilde t,\widetilde x)
      \widetilde v(\widetilde s,\widetilde y)^{\mathsf T}
    \right]\\
    &=\frac{T^2}{L^2}
      C_{\mathrm{pl}}\bigl(T(\widetilde t-\widetilde s),
      L(\widetilde x-\widetilde y)\bigr).
\end{split}
    \label{eq:rescaled-covariance}
\end{equation}

We now check how this rescaling affects the coefficients of the velocity field covariance in the Fourier space. We consider the following change of variables
\begin{equation}
    \widetilde x=\frac{x}{L},
    \qquad
    \widetilde k=Lk,
    \label{eq:rescaled-x-k}
\end{equation}
and, from this transformation, we obtain that the Fourier transform of the velocity field covariance satisfies the following relation,
\begin{align}
    \widehat{\widetilde C}_{\mathrm{pl}}
      (\widetilde t-\widetilde s,\widetilde k)
    &=\int_{\mathbb T_1^d}
      \widetilde C_{\mathrm{pl}}(\widetilde t-\widetilde s,\widetilde x)
      e^{-i\widetilde k\cdot\widetilde x}\,d\widetilde x
      \notag\\
    &=\frac1{L^d}
      \int_{\mathbb T_L^d}
      \frac{T^2}{L^2}
      C_{\mathrm{pl}} \bigl(T(\widetilde t-\widetilde s),x\bigr)
      e^{-ik\cdot x}\,dx
      \notag\\
    &=T^2L^{-2-d}D_2
      \exp\!\left(-D_3|k|^{2\beta}T
      |\widetilde t-\widetilde s|\right)
      |k|^{-d-2\alpha}P_k^\perp
      \notag\\
    &=T^2L^{-2-d}D_2
      \exp\!\left(-D_3TL^{-2\beta}
      |\widetilde k|^{2\beta}
      |\widetilde t-\widetilde s|\right)
      L^{d+2\alpha}
      |\widetilde k|^{-d-2\alpha}P_{\widetilde k}^\perp
      \notag\\
    &=T^2L^{-2+2\alpha}D_2
      \exp\!\left(-TL^{-2\beta}D_3
      |\widetilde k|^{2\beta}
      |\widetilde t-\widetilde s|\right)
      |\widetilde k|^{-d-2\alpha}P_{\widetilde k}^\perp.
    \label{eq:rescaled-fourier-covariance}
\end{align}
Hence, the rescaled velocity field retains the same covariance structure as the original model, namely \(\widetilde v\) is an Ornstein--Uhlenbeck field with parameters
\begin{equation}
    \widetilde D_2
    =T^2L^{-2+2\alpha}D_2,
    \qquad
    \widetilde D_3
    =TL^{-2\beta}D_3.
    \label{eq:rescaled-D2-D3}
\end{equation}
In particular, if we require both rescaled coefficients to be of order one, \(\widetilde D_2\approx1\) and \(\widetilde D_3\approx1\), this amounts to choosing
\begin{equation}
    D_2\approx\frac{L^{2-2\alpha}}{T^2},
    \qquad
    D_3\approx\frac{L^{2\beta}}{T}.
    \label{eq:unit-rescaled-parameters}
\end{equation}

For the parameter values
\[
    D_2=5,\qquad D_3=1,\qquad \beta=0.5,
    \qquad
    \alpha=0.2,
    \qquad
    L=20,\qquad T=10,
\]
the rescaled coefficients become
\[
    \widetilde D_2=1.21%9.32,
    \qquad
    \widetilde D_3=2,
\]
while for $T=15$ they become
\[
    \widetilde D_2=9.32,
    \qquad
    \widetilde D_3=0.75.
\]
Thus, after rescaling, the model is defined on a unit spatial domain and over a unit time interval with covariance coefficients \(\widetilde D_2,\widetilde D_3\) of order one. 

This provides a convenient normalization for interpreting the numerical simulations. However, it does not determine the characteristic spatial or temporal scale at which spontaneous particle separation, if present, may become observable.

\section*{Acknowledgment}
M.M. and S.M. acknowledge partial support by Italian Ministry of University and Research under the PRIN 2022 project `Convergence and Stability of Reaction and Interaction Network Dynamics (ConStRAINeD)' (2022XRWY7W, CUP I53D23002460001) and by Istituto Nazionale di Alta Matematica (INdAM) through the GNAMPA Project `Fluidodinamica stocastica: irregolarit\`a, trasporto e fenomeni di regolarizzazione' (CUP E53C25002010001).

The authors have used AI for proofreading and improvements in Appendix \ref{app:scales}.

%\bibliography{biblio.bib}{}
%\bibliographystyle{plain}
\bibliographystyle{abbrv}
\bibliography{biblio}

\end{document}